\documentclass[prb,twocolumn,showpacs]{revtex4-1}
\usepackage{graphicx}
\usepackage[colorlinks=true,
            linkcolor= red,
            urlcolor=blue,
            citecolor=blue]{hyperref}
\usepackage{amssymb}
\usepackage{dcolumn}
\usepackage{float}
\usepackage{bm}
\usepackage{multirow} 
\usepackage{booktabs}
\usepackage{amsmath}
\usepackage[T1]{fontenc}
\usepackage[utf8]{inputenc}
\usepackage{lmodern}
\usepackage{booktabs}
\usepackage{color}
\usepackage{makecell}
\usepackage{color}
\usepackage{subcaption}
\usepackage[export]{adjustbox}

\newcolumntype{M}[1]{>{\centering\arraybackslash}m{#1}}

\newcommand{\eps}{\varepsilon}

\newcommand{\bs}{\boldsymbol}
\DeclareMathAlphabet{\bi}{OML}{cmm}{b}{it}

\def\be{\begin{equation}}
\def\ee{\end{equation}}
\def\bearr{\begin{eqnarray}}
\def\eearr{\end{eqnarray}}
\def\la{\langle}
\def\ra{\rangle}

\begin{document}
\title{Electrical and thermoelectric transport properties of two-dimensional fermionic systems 
with $k$-cubic spin-orbit coupling}
\bigskip
\author{Alestin Mawrie, Sonu Verma and Tarun Kanti Ghosh\\
\normalsize
Department of Physics, Indian Institute of Technology-Kanpur,
Kanpur-208 016, India}
\date{\today}
 
\begin{abstract}
We investigate effect of $k$-cubic spin-orbit interaction on electrical and thermoelectric 
transport properties of two-dimensional fermionic systems. We obtain exact analytical 
expressions of the inverse relaxation time (IRT) and the Drude conductivity for long-range 
Coulomb and short-range delta scattering potentials. The IRT reveals that the scattering is 
completely suppressed along the three directions $\theta^\prime = (2n+1)\pi/3 $ with $ n=1,2,3$. 
We also obtain analytical results of the thermopower and thermal conductivity at low temperature. 
The thermoelectric transport coefficients obey the Wiedemann-Franz law, even in the presence of 
$k$-cubic Rashba spin-orbit interaction (RSOI) at low temperature. 
In the presence of quantizing magnetic field, the signature of the RSOI is revealed through the 
appearance of the beating pattern in the Shubnikov-de Haas (SdH) oscillations of thermopower and 
thermal conductivity in low magnetic field regime. The empirical formulae for the SdH oscillation 
frequencies accurately describe the locations of the beating nodes. The beating pattern 
in magnetothermoelectric measurement can be used to extract the spin-orbit coupling constant.
\end{abstract}

\pacs{74.25.Fy, 72.80.Ey, 72.10.-d}
\maketitle

\section{Introduction}
Two-dimensional (2D) systems of fermionic charge carriers (electron/hole) remain a test 
bed for spintronics related theoretical and experimental studies \cite{winkler}. 
Experimental investigations reveal the presence of strong spin-orbit coupling (SOC) associated 
with 2D electron/hole gas in $n$-/$p$-doped GaAs/AlGaAs heterojunctions \cite{bdas,Grbic}, various 
topological insulating systems \cite{TI}, strained Ge/SiGe quantum wells, 2D electron gas in transition 
metal oxide interface like SrTiO$_3$/LaTiO$_3$. 

In two-dimensional electron gas (2DEG) formed at the III-V semiconductor 
heterostructures\cite{bdas} and in various topological insulating systems \cite{TI}, 
the RSOI is linear in momentum $k$. Whereas, in p-doped III-V semiconductor heterojunctions 
such as AlGaAs/GaAs, strained-Ge/SiGe heterojunction \cite{laroche,moriya} and the 
interfaces of some transition metal oxides (e.g. SrTiO${}_3$/LaTiO${}_3$)\cite{ref101,nayak,shanavas,ani}, 
the RSOI
is found to be cubic in momentum. 
It implies that the energy level splitting due to 
$k$-cubic SOI is $2\alpha k^3$ with $\alpha $ being the spin-orbit coupling constant. 
The $k$-cubic SOI arises due to the multi-orbital effects which can be seen in the subsequent discussion. 

To describe the 2D hole gas (2DHG) in hole-doped AlGaAs/GaAs quantum well, we generally 
require a $4\times4$ Luttinger Hamiltonian\cite{luttin1,luttin2} that 
describes the hole states $\vert 3/2,\pm3/2\rangle$ and $\vert 3/2,\pm1/2\rangle$. 
At very low temperature and low density only the lowest heavy hole (HH) 
sub-bands ($\vert 3/2,\pm3/2\rangle$) are occupied. The projection of $4 \times 4$ Luttinger 
Hamiltonian onto the HH states leads to an effective $k$-cubic\cite{2dh2,2dh3} RSOI. 
On the other hand, in transition metal oxide, SrTiO${}_3$, the conduction bands originates 
from the $3d$ orbitals of Ti ion. The atomic `${\bf l} \cdot {\bf s}$' coupling splits the 
energy levels between $yz$, $zx$ and $xy$ orbitals such that the lowest energy 
states for bulk SrTiO${}_3$ will now  consist of fourfold degenerate bands, 
with total angular momentum $J =3/2$ ($m_j=\pm3/2$ and $\pm1/2$)\cite{ref115}. 
The quantum confinement in the SrTiO${}_3$/LaTiO${}_3$ further lifts this degeneracy 
such that the two-fold spin degenerate states ($|3/2,\pm3/2\rangle$) becomes 
the lowest sub-bands which is for the $xy$ orbital\cite{ref101,ref469}. 
In all these two cases, the sub-bands $|3/2,\pm3/2\rangle$ is associated with the RSOI 
which is of the form 
$H_R=\alpha(k_-^3\sigma_+-k_+^3\sigma_-)$, where $k_\pm=k_x\pm ik_y$ and 
$\sigma_\pm=\sigma_x\pm i\sigma_y$. The SOI associated with these sub-bands lift 
their two-fold spin degeneracy.

It is necessary to understand how the SOI affect the different electrical 
and magnetic transport properties. In the past few decades, several theoretical 
and experimental studies have been done on 2DHG with $k$-cubic RSOI to explore 
the effective mass,\cite{mass1, mass2, mass3} effective Lande 
g-factor,\cite{lande} spin polarization,\cite{polar} spin rotation\cite{rotation} etc. 
The experimental and theoretical analyses of the spin Hall effect are presented in Refs. 
\citep{she1, she2, she3, she4, she5, she6}. The effect of the RSOI to the Drude weight 
and optical conductivities \cite{opt1,opt2} in 2DHG has been thoroughly investigated. 
In the presence of a magnetic field normal to the 2D system, the $k$-cubic 
SOI of the charge carriers give rise to many noble phenomena such as 
weak anti-localization \cite{wal,nayak} effect in low-field magnetoresistance \cite{Grbic}  
and beating pattern at moderate magnetic field in various magnetotransport 
coefficients\cite{Grbic,AlesGhosh}.

The study of thermoelectricity is important since it directly yields the sign of 
the dominant charge carriers and provides information about the intrinsic charge 
carriers \cite{bikaner,matveev,hone,small}. There have been detailed studies of the 
thermoelectricity in conventional 2DEG \cite{smreka,Oji}. In Ref. \cite{firoz1}, a detailed 
theoretical study of the thermoelectricity in 2DEG with linear RSOI is given. The application 
of a magnetic field normal to the plane of the 2DEG/2DHG exhibits the Shubnikov–de Haas (SdH) 
oscillations in various thermoeletrical coefficients. The effect of RSOI on the oscillations 
of the thermopower and thermal conductivity has been well studied in the 2DEG \cite{wang,firoz1}. 
As a result of RSOI, they show beating pattern with the variation of magnetic field or Fermi energy. 
These beating patterns are due to the unequal spacing of Landau levels (LL's) induced by the RSOI. 
The beating pattern of these oscillations gives a direct quantitative measurement 
of the strength of the RSOI.

In this work, we study the behavior of the electrical and thermoelectric 
transport properties of the 2D fermionic system with $k$-cubic RSOI. We study the energy 
dependence of inverse relaxation time (IRT) for two types of impurity potentials, Coulomb impurity 
potential and short-range impurity potential. These results are further used for investigating 
the carrier density and RSOI dependence of the Drude conductivity, thermopower and thermal conductivity.
For non-zero magnetic field case, we study the effect of the RSOI on the various thermoelectric 
coefficients. The effect of the RSOI is reflected by the appearance of the beating pattern 
in thermoelectric coefficients at low magnetic field. The frequencies of SdH oscillations 
and the location of the beating nodes match very well with the empirical formulae given 
in Ref. \cite{AlesGhosh}.

This paper is organized as follows. In section \ref{Sec2}, the results of the 
inverse relaxation time, Drude conductivity, thermopower and thermal conductivity 
in zero magnetic field are presented. The results of the thermopower and 
thermal conductivity in the presence of quantizing magnetic field are given 
in section \ref{Sec3}. The conclusion of this paper is given in section \ref{Sec4}.

\section{2D fermionic system in zero magnetic field}\label{Sec2}
The Hamiltonian describing the 2D fermionic system with $k$-cubic RSOI 
is given by
\begin{eqnarray}\label{hamil}
H = \frac {{\bf p}^2} { 2 m^\ast } & + & \frac { i \alpha } { 2 \hbar^3 } \big (
\sigma_{+} p_{-}^3
 - \sigma_{-} p_{+}^{3} \big).
 \end{eqnarray}
Here, $m^\ast$ is the effective mass of a fermion, 
$p_\pm=p_x\pm i p_y$, $\sigma_\pm=\sigma_x\pm i \sigma_y$, with $\sigma_{x,y}$ 
being the Pauli's matrices and $\alpha$ measures the strength of the RSOI.
The spin-split energy levels and their corresponding eigenstates are respectively 
given by
\begin{eqnarray}\eps_{\textbf{k} }^{\lambda} = \frac{(\hbar k)^2}{2m^\ast} + \lambda \alpha k^3,
\end{eqnarray}
 and 
\begin{eqnarray}
\psi_{\bf k}^\lambda({\bf r}) = \frac{e^{i \textbf{k}\cdot\textbf{r}}}{\sqrt{2\Omega}} 
\phi_{\bf k}^{\lambda}.
\end{eqnarray}
Here, ${\bf k} \equiv (k_x,k_y)$, $\lambda=\pm$ denotes two spin-split energy levels, 
$\Omega$ is the surface area of the 2D sample and 
$ \phi_{\bf k}^{\lambda} = |{\bf k},\lambda \ra =
\begin{pmatrix}
1, &
-\lambda i e^{i 3\theta}
\end{pmatrix}^\mathcal{T},$ is the spinor part of the wave function. 
Here, $\theta$ is the polar angle of the wave vector 
${\bf k}$ and $\mathcal{T}$ denotes the transpose.

The velocity operator is given by 
$ \hat {\bf v} = \partial H/\partial {\bf p}$. 
The carrier velocity in each branch is given by
$ {\bf v}^{\lambda}({\bf k}) =  \la {\bf k},\lambda| \hat{\bf v} |{\bf k},\lambda \ra = 
(\hbar k/m^* + \lambda 3 \alpha k^2/\hbar) \hat k$.  
The density of states of two spin-split energy levels are 
\begin{eqnarray}
D_\lambda(\eps) 
=\frac{D_0}{|1+\lambda 3 k_\lambda(\eps)/k_\alpha|},
\end{eqnarray}
where $D_0=m^\ast/2\pi \hbar^2$, $k_\alpha=\hbar^2/m^\ast\alpha$ and 
$k_\lambda(\eps)$ is the root of $\hbar^2k^2/2m^\ast+\lambda \alpha k^3-\eps=0$.
Note that $ D_{-}(\eps) > D_{+}(\eps) $ because $ \eps_{\bf k}^+ $ is more steeper
than $ \eps_{\bf k}^- $.
For a given carrier density $n_c$, the exact expressions of the Fermi wave vectors  
and the Fermi energy are given in Ref. \cite{2dh2} and can be rewritten
in a more compact form as 
\begin{eqnarray}\label{kF}
\tilde{k}^{\lambda}_{f} & = &  
\Big[ \big(\sqrt{1 - 16 \pi n_c l_\alpha^2} - 1 \big)/8 + 3 \pi n_c l_\alpha^2 \Big]^{1/2}
\nonumber\\ 
& - & \lambda \big (1 - \sqrt{1 - 16 \pi n_c l_\alpha^2} \big)/4
\end{eqnarray} 
and 
\begin{eqnarray}
\eps_f=\frac{2 \eps_f^0 (1-16\pi n_c l_\alpha^2)}{(1+\sqrt{1-16\pi n_c l_\alpha^2})},
\end{eqnarray}
respectively. Here, $\tilde{k}^{\lambda}_{f}=k^{\lambda}_{f}/k_\alpha$ is the 
dimensionless Fermi wave vector, $ l_\alpha=1/k_\alpha $ and
$\eps_f^0 = \pi \hbar^2 n_c/m^*$ is the Fermi energy of charge carriers in 
a conventional 2D system.

For realistic samples, $n_c \leq 10^{17} $ m${}^{-2}$ and
$\alpha \leq 0.2 $ eV-nm$ {}^3$, one can easily check that 
$ 16 \pi n_c l_{\alpha}^2 \leq 0.06 \ll 1 $. 
In this limit, we have simple expressions of $k_{f}^{\lambda}$ 
and $\eps_f$ as given by
$$ 
k_{f}^{\lambda} \simeq \sqrt{3\pi n_c} - \lambda 2\pi n_c  l_{\alpha}^2 
$$
and 
$$
\eps_f \simeq \eps_{f}^0(1 - 12 \pi n_c l_{\alpha}^2).
$$ 
The Fermi energy is reduced due to the presence of the spin-orbit coupling.

\subsection{Drude conductivity and inverse relaxation time}
Applying a weak DC electric field (${\bf E} = E_x \hat i$) along $x$-direction,
the longitudinal current density becomes $ J_x = \sigma_{xx} E_x $ with 
$\sigma_{xx}$ being the Drude conductivity.
Using the semiclassical Boltzmann transport theory, the Drude conductivity \cite{mermin} 
at low temperature is given by
\begin{eqnarray}
\sigma_{xx} = \frac{e^2}{(2 \pi)^2} \sum_{\lambda} 
\int d^2 k [v_{x}^{\lambda}({\bf k})]^{2} \tau_{\lambda}(\textbf{k})
 \Big[ - \frac{ \partial f^{0}(\eps_{\bf k})}{\partial \eps_{\bf k} } \Big].
\end{eqnarray}
Here, $\langle \hat v_x ({\bf k} )\rangle_{\lambda} = 
(\hbar k/m^\ast+\lambda 3 \alpha k^2 /\hbar)\cos\theta$ ,
$\tau_\lambda({\bf k})$ is the momentum relaxation time and
$f_{\bf k}^0 = [1+ e^{\beta(\eps_{\bf k}^{\lambda} - \eta)}]^{-1}$ is the
equilibrium Fermi distribution function with $\eta $ being the chemical
potential and $\beta = (k_BT)^{-1}$.

After performing the integrals, the Drude conductivity at $T=0$ is given by
\begin{eqnarray}\label{DrudeS}
\sigma_{xx}  & = &  \frac{e^2} {2\pi m^\ast}
\sum_\lambda\tau_\lambda(k_f^\lambda) \, { (k_f^\lambda})^2(1+\lambda 3\tilde{k}_f^\lambda),
\end{eqnarray}
where $ \tau_{\lambda}(k_f)$ is the relaxation time evaluated at the Fermi contour.

Within the relaxation time approximation,
the IRT [$1/\tau_\lambda({\bf k})$] for different energy levels ($\lambda$) 
is given by \cite{mermin}
\begin{eqnarray}
\frac{1}{\tau_{\lambda}(\epsilon_{\bf k})} = \int \frac{d^2 k^{\prime}} {(2 \pi)^2}
(1-\cos\theta^\prime)
W_{\textbf{k},\textbf{k}^{\prime}}^\lambda ,
\end{eqnarray}
where $\theta^\prime$ is the angle between the wave-vectors 
${\bf k}$ and ${\bf k}^\prime$ and $W_{\textbf{k},\textbf{k}^{\prime}}^\lambda$ is 
the intra-band transition probability of a
charge carrier from the initial state $ \textbf{k} $ to the scattered state 
$\textbf{k}^{\prime} $ is given by
\begin{eqnarray}
W_{\textbf{k},\textbf{k}^{\prime}}^{\lambda}= \frac{2 \pi N_{i}}{\hbar} 
| \langle \textbf{k}, \lambda | U_\lambda(\textbf{r}) |
\textbf{k}^{\prime},\lambda \rangle |^2 
\delta( \eps_{\bf k}^{\lambda} - \eps_{ {\bf k}^{\prime}}^{\lambda}).
\end{eqnarray} 
Here $N_i$ is the number of static impurities present in the system and 
$U_\lambda(\textbf{r})$ is circularly symmetric charge-impurity interaction 
potential. The delta function ensures the conservation of energy during the 
scattering process. The quantity 
$|\langle{\bf k},\lambda|U_\lambda(\textbf{r})|{\bf k}^\prime, \lambda \rangle|^2 = 
|U_\lambda({\bf q})|^2 F({\bf k}, {\bf k}^{\prime}) $, 
where $U_\lambda ({\bf q}) = (2\pi)^{-1} \int d^2r \, e^{i {\bf q} \cdot {\bf r} } 
U_\lambda(\textbf{r}) $ is the 
impurity potential in ${\bf k} $ space with ${\bf q} = {\bf k}-{\bf k}^\prime $ being 
the change in the wave vectors and 
$ F({\bf k}, {\bf k}^{\prime}) = |\phi_\lambda^\dagger({\bf k}^{\prime})\phi_\lambda({\bf k})|^2 $ is
the modulus square of the overlap of the spinors, which is given by
\begin{eqnarray}\label{form1}
F({\bf k}, {\bf k}^{\prime})=
|\phi_\lambda^\dagger({\bf k}^\prime) \phi_\lambda({\bf k})|^2 = 
\frac{1+\cos3\theta^\prime}{2}.
\end{eqnarray}
Equation (\ref{form1}) shows that the scattering is completely suppressed along the 
three directions $\theta^\prime= (2n+1)\pi/3$, with $n=0,1 $ and $2$, irrespective of 
the type of impurity scattering potential. 
It is interesting to compare this result with that of the heavy holes in p-doped
bulk III-V semiconductors, where only the back scattering is suppressed \cite{halder}.
For 2DHG, the complete suppression of scattering along two more directions 
($\theta^\prime = \pi/3, 5 \pi/3$)
is associated with the $k$-cubic nature of the spin-orbit coupling.  

In realistic samples, the charge carriers scatter by different nature
of impurities. 
Here we consider two different types of impurity potential, namely,
long-range screened Coulomb potential i. e. Yukawa-type potential and 
short-range ($\delta$-scatterer) scattering potential. 
We are neglecting phonon interaction with the charge carriers since we
are restricted to study in the low temperature regime $ T< 1 $ K which is 
much less than the Bloch-Gruinessen temperature $ T_{\rm BG} \sim 5-10$ K.
In presence of both the independent scatterers, one can use 
Matthiessen's rule, $ 1/\tau_{\rm tot} = 1/\tau_c + 1/\tau_s$, to compute the total
IRT ($1/\tau_{\rm tot}$).  

{\bf Yukawa-type impurity potential:}
First, we consider long-range screened Coulomb potential as
$ U_\lambda(\textbf{r}) = U_0 e^{-k_{s}^{\lambda}r}/ r $, 
where $ k_{s}^{\lambda} $ is the Thomas-Fermi (TF) 
screening wave vector and $U_{0}=Z e^2/4\pi\epsilon$ with $Ze$ being ionic charge, 
$\epsilon$ being the dielectric constant of the system. The TF screening wave vectors 
can be calculated using the standard relation 
$k_{s}^{\lambda} = Ze^2 D_{\lambda}(\eps_{f})/ (2\epsilon)$ and are given
by $ k_{s}^{\lambda} = [a_B | 1+ \lambda 3 \tilde k_{f}^{\lambda} |]^{-1}$ with 
$ a_B = 4 \pi \epsilon \hbar^2/(Ze^2 m^*) $ being the effective Bohr radius. 
The Fourier transform of this 
potential is $U_\lambda({\bf q})=2\pi U_0/\sqrt{(k_s^\lambda)^2+q^2}$
with $q^2 = 4 k^2 \sin^2\theta^\prime/2$ and 
now $W_{\textbf{k},\textbf{k}^{\prime}}^{\lambda}$ is given by
\begin{eqnarray}
W_{\textbf{k},\textbf{k}^{\prime}}^{\lambda}=
 \frac{4 \pi^3 n_{i} U_{0}^{2} } {\hbar }
\frac{ (1 +\cos 3 \theta^\prime) }{ (k_{s}^{\lambda})^2 + q^2 } 
\delta( \eps_{\textbf{k}}^{\lambda} -
\eps_{\textbf{k}^{\prime}}^{\lambda}).\nonumber
\end{eqnarray}
Here, $ n_i = N_i/\Omega $ is the impurity density.
On further simplifications, the IRT for different energy levels can be 
written as
\begin{eqnarray}\label{tau2D}
\frac{\tau_0^c}{\tau_{\lambda}(k)}
=\frac{D_\lambda(k)} {D_0}\int_{0}^{2 \pi }
\frac{( 1 - \cos\theta^\prime)( 1 + \cos3 \theta^\prime) 
d\theta^{\prime}}{({k_{s}^{\lambda}}/{k_B})^2  
+ 4(k/{k_B})^2\sin^2(\theta^\prime/2)}, 
\end{eqnarray}
where $\tau_0^c = m^\ast/\pi\hbar n_i$ and $k_B = 1/a_B$. 
The IRT is thus directly proportional to the density of states $D_{\lambda} (k) $ 
and the impurity density as expected.
After performing the angular integral, we get exact analytical expressions of 
$\tau_\lambda(k)$ as given by
\begin{eqnarray} \label{tau_exact}
\frac{\tau_{0}^c}{\tau_\lambda(k)} & = & \frac{D_\lambda(k)}{2D_0} 
\frac{\pi k_B^2}{k^8}\Bigg[2k^6 + 3k^4{(k_s^\lambda})^2 + 
4k^2{(k_s^\lambda})^4 + (k_s^\lambda)^6  \nonumber \\
& - & \sqrt{4 k^2 + (k_s^{\lambda})^2 } \big\{ k^4 + 2 (k k_s^{\lambda})^2 + 
(k_s^\lambda)^4 \big\} k_s^{\lambda} \Bigg].
\end{eqnarray}
In the limit of small $ k $, 
\begin{eqnarray*}
\frac{ \tau_{0}^c}{\tau_\lambda(k)} \simeq 2\pi \frac{(k_B/k_s^\lambda)^2}{|1+\lambda 3k/k_\alpha|}
[1-3(k/{k_s^\lambda})^2].
\end{eqnarray*} 
The exact expression of the Drude conductivity at very low temperature
can easily be obtained using Eqs. (\ref{tau_exact}) and (\ref{kF}) into Eq. (\ref{DrudeS}).

\begin{figure}[!htbp]
\begin{center}\leavevmode
\centering\captionsetup{justification=RaggedRight}
\includegraphics[width=89mm,height=66mm]{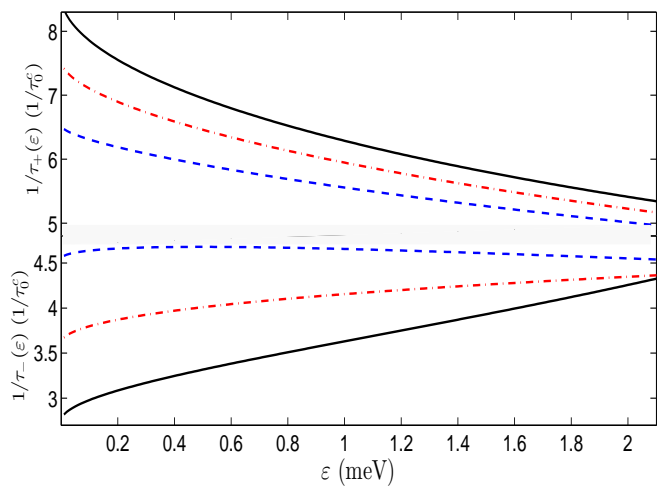}
\caption{Plots of $1/\tau_{\lambda} $ vs energy $E$ for long-range impurity potential for 
different values of $\alpha$: $\alpha = \alpha_0=0.05$ eV nm$^3$ (dashed-blue), 
$\alpha = 2\alpha_0$ (dotted-dashed-red) and $\alpha = 3\alpha_0$ (solid-black).
 }
\label{tauLR}
\end{center}
\end{figure}

\begin{figure}[!htbp]
\centering\captionsetup{justification=RaggedRight}
\begin{center}\leavevmode
\includegraphics[width=87mm,height=65mm]{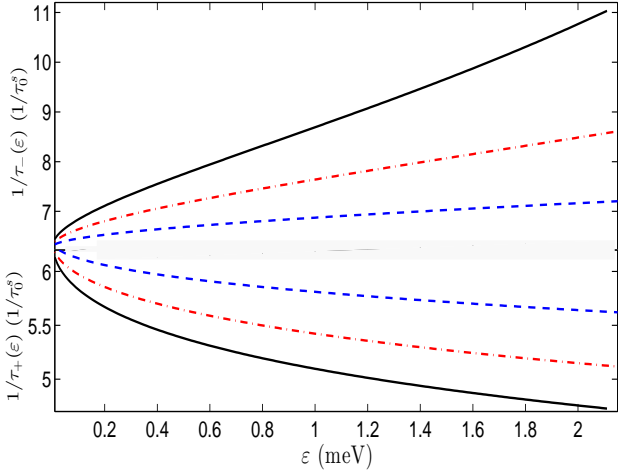}
\caption{Plots of IRT as a function of energy for delta scatterer for three different values of 
$\alpha$: $\alpha=\alpha_0$ (dashed-blue), $\alpha = 2\alpha_0$ (dotted-dashed-red) and  
$\alpha = 3\alpha_0$ (solid-black).}
\label{tauSR}
\end{center}
\end{figure}

Let us consider TF limit ($k_s^\lambda\gg k_f^\lambda$)
before presenting the numerical results of the IRT and Drude conductivity. 
In the TF limit, Eq. (\ref{tau2D}) is simplified to 
\begin{eqnarray}
\frac{\tau_0^c}{\tau_{\lambda}(k)}
& = & 2\pi  \frac{| 1 + \lambda 3 k_{f}^{\lambda}/k_{\alpha}|^2}{| 1+ \lambda 3 k/k_{\alpha}|}.
\end{eqnarray}
On the other hand, Eq.  (\ref{DrudeS}) in the TF limit is simplified as 
$ \sigma_{xx}= 2\pi n_c e^2 \tau/m^*$, independent of $\alpha$.

It would be interesting to compare this result with that of conventional 2DEG 
with $k$-linear RSOI $ H_{\rm so} = \alpha_1 (\sigma_+ p_- - \sigma_- p_+)/(2\hbar)$. 
Here, $\alpha_1$ is the strength of RSOI. Following the similar steps,
we obtain 
\begin{eqnarray}
\frac{\tau_0^c}{\tau_\lambda(k)} & = & 2 \Big(\frac{k_B}{k_{s}^{\lambda} }\Big)^2
\frac{k}{|k+ \lambda k_{\alpha_1}|}
\int_{0}^{2\pi}
\frac{(1-\cos\theta^\prime) \cos^2(\theta^{\prime}/2)}{1+4(k/k_s^\lambda)^2\sin^2\theta^\prime/2} 
\nonumber\\
& = & 2\pi \Big(\frac{k_Bk_s^\lambda}{2k^2}\Big)^2  \frac{k}{|k+\lambda k_{\alpha_1}|} 
\Big[ 1 + 2 \Big(\frac{k}{k_s^\lambda}\Big)^2 - \sqrt{1 + \Big(\frac{2k}{k_s^\lambda} \Big)^2} \Big],\nonumber
\end{eqnarray}
where $k_{\alpha_1}=m\alpha_1/\hbar^2$.
In the limit of small $k$, 
\begin{eqnarray*}
\frac{\tau_0^c}{\tau_\lambda(k)}\simeq \pi \Big( \frac{k_B}{k_s^\lambda} \Big)^2 
\frac{k}{|k+\lambda k_{\alpha_1}|}  \Big[1 - 2 \Big(\frac{k}{k_s^\lambda} \Big)^2 \Big].
\end{eqnarray*}
Note that the inverse relaxation time goes to zero at $k=0$ for linear RSOI case 
but remains finite for cubic RSOI case.

\begin{figure}[!htbp]
\centering\captionsetup{justification=RaggedRight}
\begin{center}\leavevmode
\includegraphics[width=88mm,height=60mm]{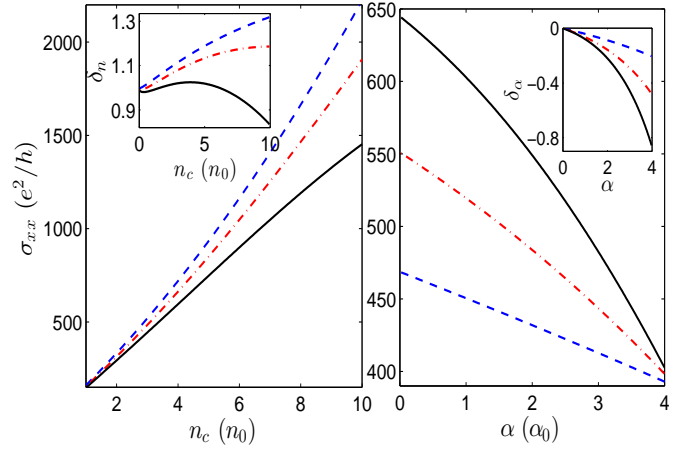}
\caption{Plots of the Drude conductivity for long-range Coulomb impurity potential. Left panel: 
$\sigma_{xx}$ vs $n_c$ (in units of $n_0=1\times10^{15}$ m$^{-2}$) for 
$\alpha=\alpha_0$ (dashed-blue), $\alpha=2\alpha_0$ (dotted-dashed-red) and 
$\alpha=3\alpha_0$ (solid-black). Right panel: $\sigma_{xx}$ vs $\alpha$ for $n_c=5n_0$ (solid-black), 
$n_c=3n_0$ (dotted-dashed-red) and $n_c=n_0$ (dashed-blue).
The upper insets show the exponent $\delta_n $ and $\delta_\alpha$ .}
\label{sigLR}
\end{center}
\end{figure}

\textbf{Short-range disorder}: Now we consider
short-range impurity potential for disorder taken as 
$U(\textbf{r})= U_s \sum_i \delta(\textbf{r}-\textbf{r}_i)$, where $U_s$ is 
the strength of the zero-range impurity potential 
having dimension of energy times area and $\textbf{r}_i$ is the position 
vector of the $i$-th impurity. 
Here the Fourier transformation of this potential is simply $U({\bf q})=U_s$ and 
\begin{eqnarray}
W_{\textbf{k},\textbf{k}^{\prime}}^{\lambda}
&=&
\frac{\pi n_{i} U_{s}^{2} } {\hbar }
( 1 +\cos 3 \theta^\prime) 
\delta( \eps_{\textbf{k}}^{\lambda} -
\eps_{\textbf{k}^{\prime}}^{\lambda} ).\nonumber
\end{eqnarray}
On further simplifications, we are able to get exact expressions of IRT for 
different branches as given by
\begin{eqnarray}\label{tau2DG}
\frac{\tau_0^s}{\tau_{\lambda}(k)}
& = &  \frac{2\pi}{|1 + \lambda 3 k/k_{\alpha}|},
\end{eqnarray}
where $\tau_0^s={4\pi \hbar^3/n_i U_s^2 m^\ast}$. 

For linear RSOI the IRT is obtained as
\begin{eqnarray}
\frac{\tau_0^s}{\tau_{\lambda}(k)}
& = &  \frac{\pi k}{|k + \lambda k_{\alpha_1}|},
\end{eqnarray}
The exact analytical expression of $\tau_{\lambda}(k) $ in Eq. (\ref{tau2DG}) 
enables us to get an exact compact expression of the Drude conductivity 
for short-range impurity potential as given by
\begin{eqnarray} \label{Drude-short}
\sigma_{xx} & = & \frac{e^2}{h} \frac{n_i}{n_c} 
\Big( \frac{2 \hbar^2}{U_s m^*} \Big)^2 
\Big[ 11 - 72 \pi n_c l_{\alpha}^2 - 6\sqrt{1 - 16\pi n_c l_{\alpha}^2} \nonumber\\
& + & \frac{3}{8 \pi n_c l_{\alpha}^2}\sqrt{1 - 16\pi n_c l_{\alpha}^2} \Big].
\end{eqnarray}
For any realistic system, $ 16 \pi n_c l_{\alpha}^2 \ll 1 $ and then the 
conductivity takes the form as 
$$ 
\sigma_{xx} \simeq \frac{e^2}{h} \frac{n_i}{n_c}
\Big( \frac{2 \hbar^2}{U_s m^*} \Big)^2
 \Big[\frac{3}{8\pi n_cl_\alpha^2} + 2 - 36 \pi n_c l_\alpha^2 \Big].
$$

{\bf Numerical Results and Discussion}: Here we present plots of $1/\tau_\lambda $ and $\sigma_{xx} $ 
given by Eqs. (\ref{DrudeS},\ref{tau_exact},\ref{tau2DG} and \ref{Drude-short}). 
For numerical calculations, we restrict ourselves to 2DHG and set $m^* = 0.41m_0$ 
with $m_0$ being the free electron mass, $\epsilon = 12.8$, $ n_i = 2 \times 10^{12}$ m${}^{-2}$ 
and we will scale the RSOI strength in units of $ \alpha_0$. Note that the system is far 
away from the TF limit for the chosen parameters here.  

We present IRT versus energy for three different values of $\alpha$ in Fig. (\ref{tauLR}) 
and Fig. (\ref{tauSR}) for long-range and short range impurity potentials, respectively. 
It is interesting to note that the difference $ \Delta(1/\tau(\eps)) = 1/\tau_-(\eps) - 1/\tau_{+}(\eps) $ 
decreases with the increase of energy ($\varepsilon$) for long-range impurity potential 
and increases with $\varepsilon$ for short-range scattering potential.

\begin{figure}[!htbp]
\centering\captionsetup{justification=RaggedRight}
\begin{center}\leavevmode 
\includegraphics[width=88mm,height=55mm]{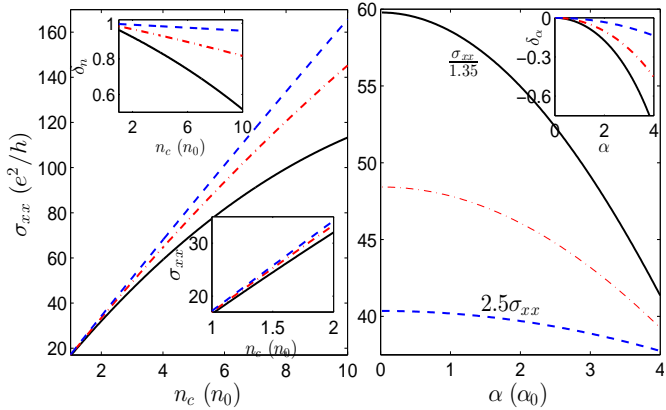}
\caption{Plots of the Drude conductivity for short-range impurity potential. 
Left panel: $\sigma_{xx}$ vs $n_c$ for $\alpha=\alpha_0$ (dashed-blue), 
$\alpha=2\alpha_0$ (dotted-dashed-red) and $\alpha=3\alpha_0$ (solid-black).
Right panel: $\sigma_{xx}$ vs $\alpha$ for $n_c = 5n_0$ (solid-black), 
$n_c=3n_0$ (dotted-dashed-red) and $n_c=n_0$ (daashed-blue). We have scaled down 
$\sigma_{xx}$ as $\sigma_{xx}/1.35$ (solid-black) and scaled up $\sigma_{xx} $ as 
$2.5\sigma_{xx}$ (dashed-blue).
The upper insets show the exponent $\delta_n $ and $\delta_\alpha$.}
\label{sigSR11}
\end{center}
\end{figure}

In Fig. (\ref{sigLR}) and Fig. (\ref{sigSR11}), we present the variations of the 
Drude conductivity with $n_c$ and $\alpha$ for long-range and short-range scattering 
potentials, respectively. In both the cases, the Drude conductivity decreases with 
increasing $\alpha$ and increases with increasing carrier density. The effective 
exponents ($\delta_n, \delta_{\alpha}$) of the carrier density and $\alpha $ dependence 
of the Drude conductivity can be obtained from the relation 
$\delta_{q}= d \log \sigma_{xx}/ d\log q$, where $q=n_c,\rm{ }\alpha$. 
The exponent $\delta_n $ versus $n_c$ and $\delta_\alpha $ versus $\alpha $ are 
shown in the upper inset of Fig. (\ref{sigLR}) and Fig. (\ref{sigSR11}), respectively. 
The density dependence of the Drude conductivity is no longer linear for both type of 
impurity potentials since $\delta_n $ deviates from unity. For long-range potential and 
large density, $\delta_n <1 $. On the other hand, $\delta_n $ is always less than one for 
short-range impurity potential. The exponent $\delta_\alpha $ is always negative for 
both type of potentials.

\subsection{Thermoelectric coefficients}\label{sec4}

In the linear response regime, electric current density (${\bf J}^e$) and 
thermal current density (${\bf U}$) in response to the combined effects 
of uniform electric field ${\bf E}$ and temperature gradient ${\boldsymbol\nabla} T$ 
are written as
\begin{eqnarray}\label{currentD}
{\bf J} = \mathcal{L}^{11} {\bf E} + \mathcal{L}^{12}(-{\boldsymbol\nabla} T)
\end{eqnarray}
and
\begin{eqnarray}
{\bf U} = \mathcal{L}^{21} {\bf E}  + \mathcal{L}^{22}(-{\boldsymbol\nabla} T).
\end{eqnarray}
Here, the transport coefficients are 
$\mathcal{L}^{11}=\mathcal{L}^{(0)}$, 
$\mathcal{L}^{12}=-\mathcal{L}^{(1)}/eT$, 
$\mathcal{L}^{21}=-\mathcal{L}^{(1)}/{e}$ 
and $\mathcal{L}^{22}=\mathcal{L}^{(2)}/{e^2T}$, 
where the different integrals $\mathcal{L}^{(r)}$ are defined as
\begin{eqnarray}\label{exactly}
&{}\mathcal{L}^{(r)} =\frac{e^2}{4\pi^2} \sum_{\lambda}^{} 
\int d^2k \, \tau_\lambda({\bf k}) \, 
{{\bf v}_{\bf k}^{\lambda}}^2  (\eps_{\bf k}^{\lambda} - \eta)^r 
\Big( - \frac{\partial f_{\bf k}^0}{\partial \eps_{\bf k}^{\lambda} } \Big).
\end{eqnarray}

The thermopower (Seebeck coefficient) is defined as 
$ S = {\bs \nabla} V/{\bs \nabla} T $, where $ {\bs \nabla} V $
is the voltage gradient developed due to the thermal gradient ${\bs \nabla} T $.  
Under an open circuit condition, the thermopower (Seebeck coefficient) and 
thermal conductivity are defined as 
$ S = (\mathcal{L}^{11})^{-1} \mathcal{L}^{12}$ and 
$ \kappa = \mathcal{L}^{22}-\mathcal{L}^{21} S $,
respectively.

Assuming ${\bf E} = E_x \hat i $ and $ {\bs \nabla T} = (\nabla_x T) \hat i$,
we need to evaluate $x$-component of the transport coefficients 
$(\mathcal{L}^{ij} $  with $i,j=1,2 $) 
for evaluation of thermopower and thermal conductivity.
In the low-temperature limit ($ k_B T \ll \eps_f $), 
the transport coefficients are obtained as
\begin{eqnarray}\label{condxx}
\mathcal{L}_{xx}^{11}& = & \frac{e^2 }{4\pi \hbar^2} 
\sum_\lambda  \tau_{\lambda}(\eps_f) \Big[ 2\eps_f+\lambda\alpha (k_f^\lambda)^3 \Big] \\
\mathcal{L}_{xx}^{12} & = & \frac{\pi}{12} \bigg(\frac{e k_BT}{\hbar}\bigg)^2 
\sum_\lambda \tau_{\lambda}(\eps_f) \bigg[ \frac{2+\lambda 9 \tilde{k}_f^\lambda}
{1+\lambda 3 \tilde{k}_f^\lambda}\bigg] \label{k1xx} \\
\mathcal{L}_{xx}^{22}& = & \frac{1}{12\pi} \Big( \frac{\pi k_B}{\hbar}\Big)^2 T 
\sum_\lambda \tau_\lambda(\eps_f) 
\Big[2 \epsilon_f + \lambda \alpha (k_f^\lambda)^3\Big] \label{loren} \\
&=&\frac{1}{3}\Big(\frac{\pi k_B}{e}\Big)^2T\mathcal{L}_{xx}^{11}.\nonumber
\end{eqnarray}
Here, we consider the relaxation time $\tau_{\lambda}(\eps_f) $ calculated for 
either Coulomb-type or short-range impurity potential or both the scattering potentials. 
We immediately see the ratio of transport coefficients obey  the 
Wiedemann-Franz law,  $\mathcal{L}_{xx}^{22}/T\mathcal{L}_{xx}^{11} 
= (\pi k_B)^2/(3e^2) $, at very low temperature.
Therefore, the Wiedemann-Franz law holds even in the presence of $k$-cubic RSOI.

\begin{figure}[!htbp]
\centering\captionsetup{justification=RaggedRight} 
\begin{center}\leavevmode
\includegraphics[width=89mm,height=65mm]{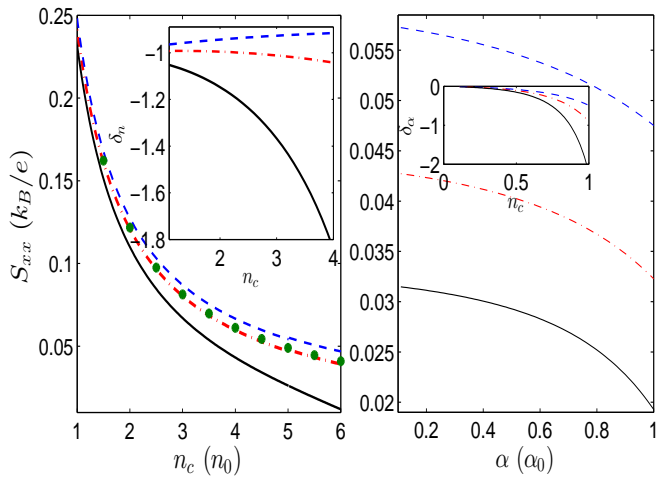}
\caption{Plots of the thermopower $S_{xx}$  for long-range Coulomb scatterer at temperature 
$T=0.5$ K. Left panel: $S_{xx}$ vs $n_c$ for $\alpha=\alpha_0$ (dashed-blue), 
$\alpha=2\alpha_0$ (dotted-dashed-red) and $\alpha=3\alpha_0$ (solid-black). 
Right panel: $S_{xx} $ vs $\alpha $ for $n_c =10n_0 $ (dashed-blue), $n_c = 7n_0$ 
(dotted-dashed-red) and $n_c = 5n_0$ (solid). 
The green dots are obtained by using Eq. (\ref{sxx}). 
The insets show the exponents $\nu_n=\frac{d\log S_{xx}}{d \log n_c}$ 
and $\nu_\alpha=\frac{d\log S_{xx}}{d \log \alpha}$.} 
\label{thermopower}
\end{center}
\end{figure}

Using Eqs. (\ref{condxx}) and (\ref{k1xx}), the thermopower 
can be written as
\begin{eqnarray}\label{sxx}
S_{xx}=\frac{\pi^2k_B}{2e}\frac{k_BT}{\eps_f}  
\frac{\sum_{\lambda}^{} \tau_\lambda(\varepsilon_f)\bigg\{1-\frac{1}{3 (1+\lambda 3\tilde{k}_f^\lambda)}\bigg\}}
{\sum_{\lambda}^{} \tau_\lambda(\varepsilon_f)\bigg\{1+\lambda\alpha (k_f^\lambda)^3/2\eps_f\bigg\}}.
\end{eqnarray}

Let us analyze the characteristics of the thermopower and the thermal conductivity numerically 
and comparing with the value obtained by the low temperature approximation. 
For numerical calculation, the value of relaxation time $\tau_{\lambda}$ is taken 
from Eq. (\ref{tau_exact}) for long-range Coulomb scattering. The variation of the exact 
values of $S_{xx}$ (using Eq. (\ref{exactly}) in the definition of $S_{xx}$) with respect 
to $n_c$ for different strength of RSOI is shown in Fig. (\ref{thermopower}). {
The exact numerical results match very well with the approximate results [Eq. (\ref{sxx})] 
shown by green dots in Fig. (\ref{thermopower})). 
Thermopower decreases with increasing strength of RSOI ($\alpha$) as shown in the right 
panel of Fig. (\ref{thermopower}).

\begin{figure}[!htbp]
\centering\captionsetup{justification=RaggedRight}
\begin{center}\leavevmode
\includegraphics[width=89mm,height=65mm]{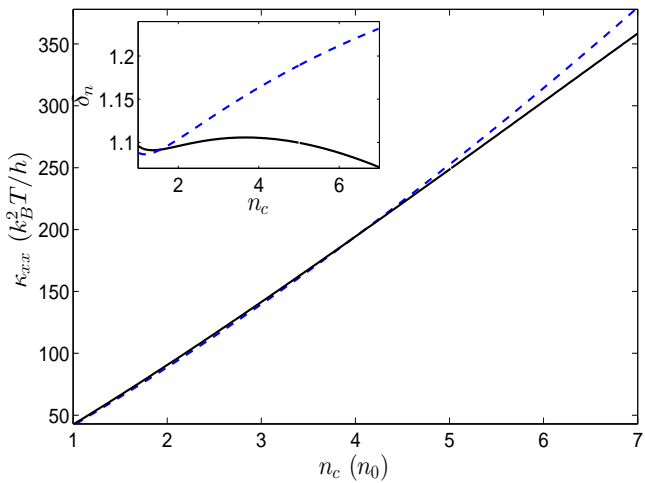}
\caption{Plots of the thermal conductivity ${\kappa_{xx}}$ vs $n_c$ for long-range Coulomb scatterer 
at temperature $T=0.5$ K for $\alpha=\alpha_0$(dashed-blue) and $\alpha=3\alpha_0$ (black). 
The inset contains the exponent $\tilde{\nu}_n=\frac{d\log \kappa_{xx}}{d\log n_c}$ }
\label{TherCon}
\end{center}
\end{figure}
The variation of the thermal conductivity with respect to $n_c$ for two different RSOI's strength 
is shown in Fig. (\ref{TherCon}). 
It is easy to see that the RSOI diminishes the thermopower only at higher carrier density. 
It is an increasing function of the carrier density and there are not much 
significant variations of the thermal conductivity with $\alpha$.

\section{2D fermions in presence of quantizing  magnetic field}\label{Sec3} 
In the presence of a quantizing magnetic field ${\bf B} = B \hat z$, the Hamiltonian in 
Eq. (\ref{hamil}) changes to\cite{Liu, Zarea}
\begin{eqnarray}\label{hamil2}
H=\frac{{\bf P}^2}{2m^\ast} + 
\frac{i\alpha}{2\hbar^3}\big(P_{-}^3\sigma_{+} - P_{+}^3 \sigma_{-} \big) - 
\frac{3}{2}g^\ast{\mu_{B}}{\boldsymbol{\sigma}} \cdot {\bf{B}}.
\end{eqnarray}
Here, ${\bf P}={\bf p}- e{\bf A}$ with ${\bf A}$ being the vector potential, 
$P_\pm = P_x \pm i P_y$, $g^\ast$  is the effective Lande-g factor and $\mu_{B}$ is 
the Bohr magneton. We choose the Landau gauge ${\bf{A}}= xB  \, \hat{y}$ so that $k_y$ is a 
good quantum number. The Landau levels in units of $\hbar \omega_c$ with $\omega_c =eB/m^*$ 
being cyclotron frequency, are given by 
\begin{equation}
\tilde{\eps}_n^{\lambda}=n-1+
\lambda\sqrt{\tilde{\eps}_{n\alpha}^2+\tilde{\eps}_0^2}.
\end{equation}
Here, $n \ge 3$, $\tilde{\eps}_0=3/2 - \chi$ with $ \chi=3g^\ast m^\ast/(4m_0)$ 
and $\tilde{\eps}_{n\alpha}=\tilde{\alpha}\sqrt{8n(n-1)(n-2)}$. 
The dimensionless parameter $\tilde{\alpha}$ is defined as $\tilde{\alpha}=l_{\alpha}/l_c$ 
with $l_c=\sqrt{\hbar/(eB)}$ is the magnetic length. The corresponding eigenstates are given by
\begin{eqnarray} 
\psi_{\xi}^+({\bf r}) = \frac{e^{ik_yy}}{\sqrt{L_y\mathcal{A}_n}} 
\begin{pmatrix} 
\phi_{n} (\tilde x )
\\ \mathcal{D}_n\phi_{n-3} (\tilde x)
\end{pmatrix}
\end{eqnarray}
and
\begin{eqnarray} 
\psi_{\xi}^-({\bf r}) = \frac{e^{ik_yy}}{\sqrt{L_y\mathcal{A}_n}} 
\begin{pmatrix} -\mathcal{D}_n\phi_{n} (\tilde x)
\\ \phi_{n-3}(\tilde x) \end{pmatrix},
\end{eqnarray}
where $\xi\equiv \{n,k_y \}$ is a set of two quantum numbers, 
$L_y$ is the system size along $y$-direction,
$ \tilde x = (x -x_c)/l_c $ with $x_c=k_yl_c^2$, $\mathcal{A}_n=1+\mathcal{D}_n^2$ with 
$\mathcal{D}_n=\tilde{\eps}_{n\alpha}/\Big(\tilde{\eps}_0 + 
\sqrt{\tilde{\eps}_0^2+\tilde{\eps}_{n\alpha}^2}\Big)$ and
$\phi_n(x)$ is the $n$-th order oscillator wave function normalized to
unity.
The first three Landau levels ($n=0,1,2$) do not get split by the RSOI. 
The Landau levels and the corresponding eigenstates for $n=0,1,2$ are given by
$ \tilde{\eps}_n=n + 1/2 - \chi $
and
\begin{eqnarray} 
\psi_{\xi}({\bf r})=\frac{e^{ik_yy}}{\sqrt{L_y}}\phi_n 
(\tilde x)\begin{pmatrix} 1
\\ 0 \end{pmatrix}.
\end{eqnarray}

\begin{figure*}[t]
\centering\captionsetup{justification=RaggedRight}
\includegraphics[width=129mm,height=66mm]{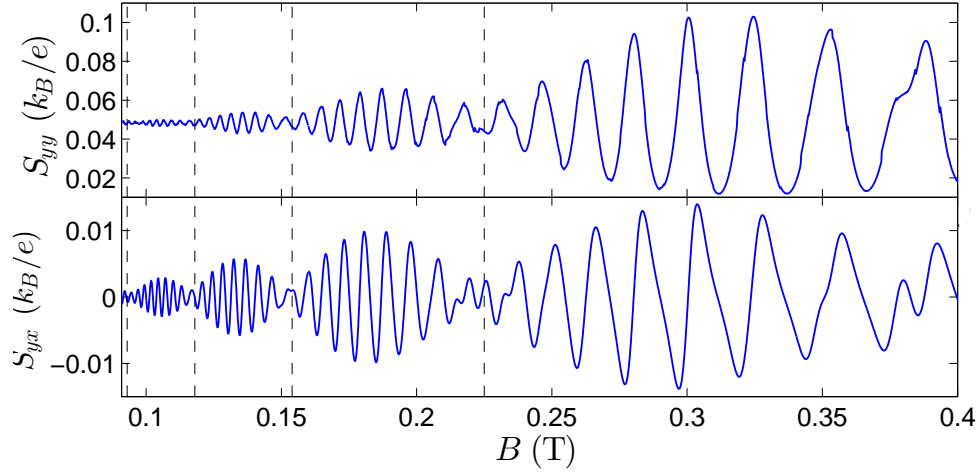}
\caption{Plots of the thermopower $S_{xx} $ and $S_{xy} $ as a function of the magnetic field $B$. 
} \label{Mag1}
\end{figure*}

\subsection{Thermoelectric coefficients}
In this section, we calculate various magnetothermoelectric coefficients. For spin-split systems, 
the various transport coefficients in presence of magnetic field are given by
\begin{eqnarray}\label{thermopowers}
S_{\mu\nu}^\lambda=\frac{1}{eT}[(\mathcal{L}^{(0),\lambda})^{-1} 
\mathcal{L}^{(1),\lambda}]_{\mu\nu},\\\label{thermalconduc}
\kappa_{\mu\nu}^\lambda=\frac{e^2}{T}[\mathcal{L}_{\mu\nu}^{(2), 
\lambda}-e T(\mathcal{L}^{(1),\lambda}S^\lambda)_{\mu\nu}],
\end{eqnarray}
\pagebreak
with
\begin{eqnarray}
\mathcal{L}_{\mu\nu}^{(r),\lambda}=\int d\eps 
\bigg[-\frac{\partial f(\eps)}{\partial \eps}\bigg](\eps - \eta)^r\sigma_{\mu\nu,\lambda}(\eps).
\end{eqnarray}
Here the indices $\mu,\nu=x,y$ and $\sigma_{\mu\nu,\lambda}(\eps)$ is the 
zero-temperature energy-dependent conductivity.
The above thermoelectric coefficients are obtained by generalizing the results 
of Refs. \cite{smreka,Oji} for the case of spin-split systems.
The total thermopower and thermal conductivities are 
$S_{\mu\nu}=\sum_\lambda S_{\mu\nu}^\lambda$ and 
$\kappa_{\mu\nu}=\sum_\lambda \kappa_{\mu\nu}^\lambda$, respectively. 
The transport of holes takes place only through a collisional mechanism because of the
zero drift velocity.

The energy-dependent zero-temperature collisional and Hall conductivities 
are given\cite{AlesGhosh} as
\begin{widetext}
\begin{equation}\label{coll}
\begin{aligned}
\sigma_{xx,\lambda}^{\rm coll}(\eps)=-\frac{e^2}{h}\frac{\Gamma}{4\pi^2}{} & 
\Big[\sum_{n<3} I_n\delta(\eps - \eps_n)\delta_{\lambda,1}
      +\sum_{n\ge3}I_n^\lambda \delta(\eps - \eps_n^\lambda)\Big],
\end{aligned}
\end{equation}
\begin{eqnarray}\label{HT}
  \sigma_{yx,\lambda}^{\rm Hall} (\eps) & = &\frac{2e^2}{h} 
\bigg[\sum_{n=0}^{1}2(n+1)\theta(\eps - \eps_n,\eps_{n+1} - \eps)\delta_{\lambda,1}
  +\bigg(\frac{P_{23}^\lambda}{{\Delta_{2,3}^\lambda}}\bigg)^2{
\theta(\eps - \eps_2,\eps_3^\lambda - \eps)}{} \nonumber\\ 
&+&2\sum_{ n\ge3}\bigg(\frac{P_n^\lambda}{{\Delta_n^\lambda}}\bigg)^2 
\theta(\eps - \eps_n^\lambda,\eps_{n+1}^\lambda - \eps)\bigg].
\end{eqnarray}
\end{widetext}

Here $I_n=2n+1$, $I_n^\lambda= [(2n-2-\lambda3)(\mathcal{D}_n^4+1) + \lambda6]/\mathcal{A}_n^2 $, 
$\theta(x,y)$ represents a double arguments Heaviside theta function which is 1 only if both 
$x$ and $y$ are positive otherwise zero,
\begin{eqnarray}
P_n^+ & = & \frac{ \sqrt{n+1}+ \mathcal{D}_n\mathcal{D}_{n+1}\sqrt{n-2}
+6\tilde \alpha \mathcal{D}_{n+1}\sqrt{2n(n-1)} }{\sqrt{2\mathcal{A}_n\mathcal{A}_{n+1}}},\nonumber\\
P_n^- & = & \frac{\mathcal{D}_n\mathcal{D}_{n+1}\sqrt{{n+1}{}}+\sqrt{{n-2}{}}-6\tilde{\alpha}
\mathcal{D}_{n+1}\sqrt{2n(n-1)}}{\sqrt{2\mathcal{A}_n\mathcal{A}_{n+1}}},\nonumber 
\end{eqnarray}

\begin{eqnarray}
P_{23}^+{}& = & ({\sqrt{3}+12\tilde{\alpha}\mathcal{D}_3})/{\sqrt{2\mathcal{A}_3}},\nonumber\\
P_{23}^-{}& = & ({\sqrt{3}\mathcal{D}_3-12\tilde{\alpha}})/{\sqrt{2\mathcal{A}_3}},\nonumber
\end{eqnarray}
\pagebreak[3]
\pagebreak[3]
\begin{eqnarray}
\Delta_{2,3}^\lambda & = & 1/2-\chi-\lambda\sqrt{(3/2-\chi)^2+48\tilde{\alpha}^2}, \nonumber \\
\Delta_n^\lambda & = & - 1 + \lambda\{\sqrt{(3/2-\chi)^2+8n(n-1)(n-2)\tilde{\alpha}^2}\nonumber \\
& - & \sqrt{(3/2-\chi)^2 + 8n(n-1)(n-2) \tilde{\alpha}^2}\}.\nonumber
\end{eqnarray}
\pagebreak[3]

Using Eqs. (\ref{coll}), (\ref{HT}), the finite temperature 
$\mathcal{L}_{yy}^{(r)}$ and $\mathcal{L}_{yx}^{(r)}$ can be written as
\begin{widetext}
\begin{eqnarray}\label{colTH}
\mathcal{L}_{yy}^{(r),\lambda} & = & \frac{e^2}{h}\frac{\Gamma}{4\pi^2} 
\Big\{\delta_{\lambda,1}\sum_{n<3} I_n (\eps - \eta)^r 
\Big[-\frac{\partial f(\eps)}{\partial \eps}\Big]_{\eps_n} + 
\sum_{n}I_n^\lambda (\eps-\eta)^r 
\Big[ - \frac{\partial f(\eps)}{\partial \eps}\Big]_{\eps_n^\lambda}\Big\}
\end{eqnarray}
and
\begin{eqnarray}\label{HalTH}
\mathcal{L}_{yx}^{(r),\lambda}&=&\frac{2e^2}{h}\bigg[\delta_{\lambda,1}
\sum_{n=0}^{1}2(n+1) \int_{\eps_n}^{\eps_{n+1}}(\eps - \eta)^r 
\Big[-\frac{\partial f(\eps)}{\partial \eps}\Big] d\eps + 
\bigg(\frac{{P_{23}^\lambda}}{{\Delta_{23}^\lambda}}\bigg)^2 
\int_{\eps_2}^{\eps_{3}^\lambda}(\eps - \eta)^r 
\Big[-\frac{\partial f(\eps)}{\partial \eps}\Big] d\eps \nonumber\\
& + & 2\sum_{n>2}\bigg(\frac{{P_n^{\lambda}}}
{{\Delta_n^\lambda}}\bigg)^2\int_{\eps_n^\lambda}^{\eps_{n+1}^\lambda}(\eps - \eta)^r 
\Big[-\frac{\partial f(\eps)}{\partial \eps}\Big] d\eps
\bigg],
\end{eqnarray}
\end{widetext}
respectively.

Using Eq. (\ref{colTH}, \ref{HalTH}) in Eq. (\ref{thermopowers}, \ref{thermalconduc}), 
we compute the various thermoelectric coefficients numerically. 
For the numerical approach we have taken the value of various parameters, 
such as carrier density $n_c=2n_0$, the effective mass $m^\ast=0.41 m_0$, 
RSOI strength $\alpha=\alpha_0$ and taking the temperature as low as $T=0.1$ K.

The beating pattern formation of the thermopower and the thermal conductivity with respect 
to the applied magnetic field are shown in Fig. (\ref{Mag1} - \ref{Mag4}). 
The beating patterns are prominent for $B\le 0.3$ T.
The beating pattern arises due to the difference in the SdH oscillation 
frequencies ($f_{\pm}$) of two spin-split branches. The beating pattern in the 
oscillation of the thermopower and thermal conductivity can be modeled as 
$\cos(2\pi f_d/B)\cos(2\pi f_a/B)$. Here $f_d=(f_{-} - f_{+})/2$ and $f_a=(f_-+f_+)/2$, 
where $f_\pm$ are the SdH oscillation frequencies.
The expression of $f_\pm$ is 
taken from Ref. \cite{AlesGhosh}, which have been used to explain the beating pattern 
observed in the SdH oscillation of the collisonal conductivity, $i.e.$
\begin{eqnarray}\label{sdh}
f_{\pm}=\frac{m^*}{\hbar e}\Bigg[\eps_{f}\mp 
 \sqrt{\frac{8\alpha^2m^{*3}{\eps_{f}}^3}{\hbar^6} + \eps_0^2}\Bigg].
\end{eqnarray}
Note that the SdH frequencies $f_\pm$ are magnetic field dependent which is coming from 
the term $\varepsilon_0$. The non-appearance of beating patterns at moderate or high magnetic field
is due to the magnetic field dependent frequencies $f_{\pm}$.
\begin{figure}[t]
\includegraphics[width=89mm,height=55mm]{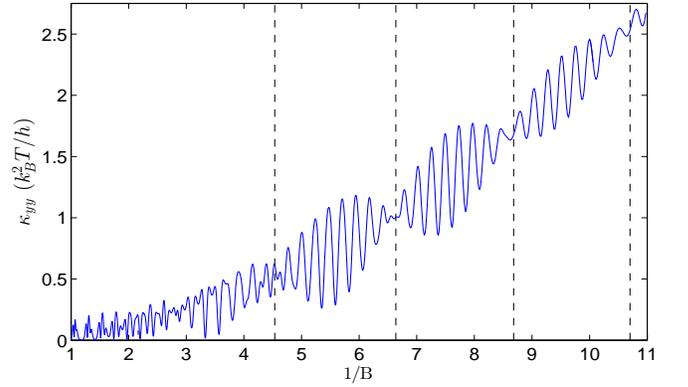}
\caption{Plots of the thermal conductivity $\kappa_{yy}$ as a function of inverse 
of magnetic field ($B$ is in Tesla).} 
\label{Mag4}
\end{figure} 
With the help of Eq. (\ref{sdh}) and by analyzing the oscillations, the location of each of 
the beating nodes are
\begin{eqnarray}\label{beat}
B_j=\phi_0 n_c\sqrt{\frac{{32\pi n_c l_\alpha^2}}{{(2j+1)^2-16 \tilde{\eps}_0^2}}}, 
\end{eqnarray}
where $\phi_0 = h/e$ is the magnetic flux quantum and $j=0,1,2,...$.
\begin{table}[ht]
\centering
\caption{Beating nodes calculated from Eq. (\ref{beat}) and that appearing in different figures.}
\begin{center}
\begin{tabular}{ |c|c|c| }
\hline
\hline
 & \multicolumn{2}{ |c| }{Position of the nodes ($B_j$)}  \\
\hline
j  &  From Fig. (\ref{Mag1} \& \ref{Mag4})  &  From Eq. (\ref{beat}) \\ \hline
2  &  0.224  &  0.225\\
3  &  0.152  &  0.154\\
4  &  0.116   &  0.118\\
5  &  0.092  &  0.093 \\
\hline
\end{tabular}
\end{center}
\end{table}
The values of $B_j's$ coincide with the nodes appearing in Fig. (\ref{Mag1} - \ref{Mag4}). 
In Fig. (\ref{Mag4}), the values of the corresponding $1/B_j's$ are marked with the vertical dashed lines.
The number of oscillations between two successive nodes
\begin{eqnarray}\label{NOSC}
N_{\rm{0}}=\frac{1}{\sqrt{32\pi n_cl_\alpha^2}}
\big[&{}\sqrt{(2j+3)^2-16\tilde{\epsilon_0}^2}\nonumber\\&{}-\sqrt{(2j+1)^2-\tilde{\epsilon_0}^2}\big].
\end{eqnarray}
To test correctness of Eq. (\ref{NOSC}), we count the number of oscillations between 
$B=0.117$ T and $B=0.152$ T, which is $N_0=8$. With the values of the different parameters 
(except $\alpha$) in Eq. (\ref{NOSC}), the value 
of RSOI becomes $\alpha=0.0529$ eV-nm$^3$ $ \approx\alpha_0$ which is the value that we have taken.

\section{Summary}\label{Sec4}
We have studied the effect of $k$-cubic Rashba spin-orbit interaction on 
electrical and magnetothermoelectric coefficients of a 2D fermions formed 
in different condensed matter systems. 
We obtained exact analytical expressions of the IRT and the Drude conductivity for long-range 
and short-range impurity potentials. Our study revealed that the scattering is completely 
blocked along three different angles $ \theta^{\prime} = (2n+1)\pi/3 $ ($n=0,1,2$), 
irrespective of the type of impurity potentials. The IRT remains finite at $k\rightarrow 0$ 
in contrast to $k$-linear RSOI case.

At zero magnetic field, we find that the RSOI diminishes the thermopower. 
However the thermal conductivity is an increasing function of the carrier density 
and there are not much significant variation with respect to the strength of RSOI. 
We have obtained analytical results of the thermopower and thermal conductivity 
at low temperature ($\eps_f\gg k_BT$) using Sommerfeld's expansion. 
The Wiedemann-Franz law remains valid even in the presence of RSOI. 

For non-zero magnetic field, the signature of the RSOI is revealed through the 
beating pattern appeared in thermopower and thermal conductivity at low magnetic field regime. 
The empirical formula of the SdH oscillation frequencies allows us to 
carefully note the location of each node in the beating pattern of the oscillations.
Magnetothermoelectric measurements can be used to determine the strength of the
$k$-cubic Rashba SOI.

\end{document}